\documentclass[twocolumn,showpacs,showkeys,preprintnumbers,amsmath,amssymb]{revtex4}

\usepackage{graphicx}
\usepackage{dcolumn}
\usepackage{bm}

\begin{document}


\preprint{ }

\title{Demonstrating quantum algorithm acceleration with NMR quantum computer}

\author{Mikio Nakahara$^{1}$, Yasushi Kondo$^{1}$, Kazuya Hata$^{1}$
and Shogo Tanimura$^{2}$\\}

\affiliation{%
$^{1}$Department of Physics, Kinki University, Higashi-Osaka 577-8502, Japan\\
$^2$Graduate School of Engineering, 
Osaka City University\\
Sumiyoshi-ku, Osaka, 558-8585, Japan\\
}%

\date{\today}

\begin{abstract}
In general, a quantum circuit is constructed with elementary gates, such as 
one-qubit gates and CNOT gates. 
It is possible, however,
to speed up the execution time of a given circuit
by merging those elementary gates together into larger modules, such that the 
desired unitary matrix
expressing the algorithm is directly implemented. We demonstrate this
experimentally
by taking the two-qubit Grover's algorithm implemented in NMR quantum 
computation,
whose pseudopure state is generated by cyclic permutations of the 
state populations. This is the first exact time-optimal
solution, to our knowledge, obtained for a self-contained quantum algorithm.
\end{abstract}

\pacs{03.67.Lx, 82.56.Jn}
\keywords{NMR quantum computer, time-optimal control, Grover's algorithm}

\maketitle

\section{Introduction}

A quantum computer is expected to solve some of computationally
hard problems for a conventional digital computer \cite{ref:1}. The
realization
of a practical quantum computer is, however, still challenging
in many respects \cite{ref:2}. One of the obstacles to any realization is
a phenomenon known as ``decoherence''. The number of gate
operations is severely limited since a quantum state is vulnerable
due to interactions with the surroundings. Many strategies
to overcome internal and external decoherence have been proposed, 
such as (1) quantum error correcting codes \cite{ref:3} (2) 
decoherence-free subspaces \cite{ref:4} (3) 
holonomic quantum computation \cite{ref:5}, among others. 
In a conventional design of a quantum circuit, the
so-called elementary set of gates \cite{ref:6, ref:a, ref:b} 
such as single-qubit $SU(2)$ rotations
and CNOT gates, are utilized. This design is motivated by the universality
theorem proved in Ref. \cite{ref:6}.
Suppose we are to implement an $n$-qubit unitary matrix
$U_{\rm target}$. In the conventional implementation this matrix is 
decomposed into a product of $SU(2)$ matrices acting bewteen a pair of
basis vectors. Then a CNOT gate transforms one of the basis vectors
to a new vector, so that the pair of the vectors forms a subspace
corresponding to a single qubit on which the $SU(2)$ matrix acts.

Quantum algorithm acceleration is a totally new approach to the
decoherence issue. This principle is originally proposed 
in the context of the holonomic quantum computation \cite{ref:7, ref:8,
ref:9} and of Josephson junction qubits \cite{ref:10, ref:11, ref:12}.
What is required to implement is not individual
elementary gates but rather the $n$-qubit matrix 
$U_{\rm target} \in U(2^n)$ realizing the given quantum algorithm.
This matrix is directly implemented by properly choosing
the control parameters in the Hamiltonian. The variational
principle tells us that the gate execution time reduces, in general,
compared to the conventional construction since the conventional gate sequence 
belongs to the possible solutions in direct implementation.

The proposal \cite{ref:10, ref:11, ref:12} has been made for 
fictitious Josephson
charge qubits, which are still beyond reach. It is the purpose
of this paper to demonstrate the acceleration of a quantum algorithm
using an NMR quantum computer at our hand. For this demonstration,
we employ two-qubit Grover's search algorithm
whose initial state $|00 \rangle$ is generated as a pseudopure state
by cyclic permutations of the state populations \cite{ref:13}. 
In this process, we need to prepare three different
initial states using $SU(4)$ transformations acting on a thermal
equilibrium ensemble. It is found that the optimized pulse sequence
reduces the gate operation time to 25\% of the conventional 
pulse sequence operation time in two ensembles while it remains unchanged
in one ensemble, which is already optimized using the conventional
pulse sequence. 

The next section is devoted to the formalism of our approach.
The Hamiltonian for the two-qubit molecule is introduced and
the time-evolution operator is defined.
Section III is the main part of this paper, where the exact solutions
for Grover's algorithm are obtained. These solutions are experimentally
verified with our NMR computer. Section IV is devoted to summary
and discussion.

\section{Time-Optimal Path in NMR Quantum Computation}

In the present paper, we are concerned with an NMR quantum computer
with a two-qubit heteronucleus molecule. To be more specific, we use 
Carbon-13 labeled chloroform
as a computational resource throughout our theoretical and experimental
analyses. The Hamiltonian of the molecule is
\begin{eqnarray}
H(\gamma) &=& -\omega_{11} \left[\cos \phi_1 (\sigma_x \otimes I_2/2) + 
\sin \phi_1 (\sigma_y \otimes I_2/2) \right]\nonumber\\
& &-\omega_{12} \left[\cos \phi_2 (I_2 \otimes \sigma_x /2) + 
\sin \phi_2 (I_2 \otimes \sigma_y/2) \right]\nonumber\\
& &+ 2\pi J \sigma_z \otimes \sigma_z/4,
\label{eq:ham}
\end{eqnarray}
in the rotating frame of each nucleus.
Here $I_2$ is the unit matrix of order 2 while $\sigma_k$ is
the $k$th Pauli matrix. The parameter $\omega_{1i}$ is the 
amplitude of the RF pulse for the $i$th spin while
$\phi_i$ is its phase. These four independent control parameters
are collectively denoted as $\gamma$. 

The time-evolution operator
\begin{equation}
U[\gamma(t)] = {\mathcal T} \exp\left[-i \int_0^T
H(\gamma(t)) dt\right]
\end{equation}
associated with the Hamiltonian 
(\ref{eq:ham}) is a functional of $\gamma(t)$, where ${\mathcal T}$ is the 
time-ordered product and we employ the natural units in which $\hbar$ is set 
to unity. Note that
$U[\gamma(t)] \in SU(4)$ since (\ref{eq:ham}) is traceless. Suppose we want 
to implement a unitary
gate $U_{\rm target}$. Then our task is to find a function $\gamma(t)$
such that
\begin{equation}
U[\gamma(t)] = U_{\rm target}.
\label{eq:inverse}
\end{equation}
The solution of this 
problem is highly nontrivial and we often resort to
numerical analysis \cite{ref:7, ref:8, ref:11, ref:12}. 
Note, moreover, that we would like
to construct time-optimal solutions among many possible solutions
of Eq.~(\ref{eq:inverse}) to fight against decoherence. It must
also be examined whether the optimal solution thus obtained has
control parameters that are experimentally accessible.

It was shown in Refs.~\cite{ref:14, ref:15} that there exists a 
geometrical picture
for the time-optimal solution in the case of NMR quantum computation
for two-qubit heteronuclear molecules. Here we summarize briefly the
relevant aspects of their idea that are required for our investigation.
The crucial observation is that the one-qubit rotation is 
carried out in a negligible time compared to that required for
two-qubit operation generated by the $J$-coupling term in Eq.~(\ref{eq:ham})
and hence time required for the one-qubit rotation
can be neglected in evaluating the
gate operation time. Then it is natural to consider the homogeneous
space $SU(4)/K$, where $K \equiv SU(2) \otimes SU(2)$,
in which two-qubit gates that
differ by one-qubit operations $U_1 \otimes U_2 \in K$ are
identified. The path in the homogeneous space is then generated by
the $J$-coupling term in Eq.~(\ref{eq:ham}). 
A remark is in order.
The Hamiltonian (\ref{eq:ham}) contains neither
$\sigma_z \otimes I_2$ nor $I_2 \otimes \sigma_z$ and it seems
impossible to consider the equivalence class based on
the whole subgroup $K$.
Note, however, that these terms are easily generated 
by $\sigma_x$ and $\sigma_y$ as
\begin{equation}
e^{-i \alpha \sigma_z/2} = e^{i \pi \sigma_y/4} e^{-i \alpha \sigma_x/2}
e^{-i \pi \sigma_y/4},
\label{eq:1qubit} 
\end{equation}
for example, and hence it makes sense to identify all the elements of
$SU(4)$, that differ with each other modulo the subgroup $K$.
With the same token, the terms $\sigma_x \otimes \sigma_x$ and 
$\sigma_y \otimes \sigma_y$ may be included in the $J$-coupling part of the 
Hamiltonian since
\begin{eqnarray}
\lefteqn{e^{-i \pi J t \sigma_x \otimes \sigma_x}
= e^{i \pi(\sigma_y \otimes I_2 + I_2 \otimes \sigma_y)/4}}\nonumber\\
& &\times e^{-i \pi J t \sigma_z \otimes \sigma_z/2}
e^{-i \pi(\sigma_y \otimes I_2 + I_2 \otimes \sigma_y)/4}
\label{eq:2qubit}
\end{eqnarray}
for example. It should be noted that the set $\{\sigma_x \otimes \sigma_x,
\sigma_y \otimes \sigma_y, \sigma_z \otimes \sigma_z\}$ comprises the
Cartan subalgebra of $SU(4)$ in a relevant basis and hence
\begin{eqnarray*}
\lefteqn{e^{-i (\alpha_1 \sigma_x \otimes \sigma_x+\alpha_2 \sigma_y \otimes \sigma_y
+ \alpha_3 \sigma_z \otimes \sigma_z)}}\nonumber\\
&=& e^{-i \alpha_1 \sigma_x \otimes \sigma_x} 
e^{-i \alpha_2 \sigma_y \otimes \sigma_y}
e^{-i \alpha_3 \sigma_z \otimes \sigma_z}.
\end{eqnarray*}

Suppose we want to find a time-optimal path $\gamma(t)$ connecting
the unit matrix $I_4$ of order four and the target matrix $U_{\rm target} 
\in SU(4)$. The prescription given above suggests that we should
find the time-optimal path in $SU(4)/K$ that
connects the equivalence classes $K$ and $U_{\rm target} K$.
In other words, we have to solve the matrix equation
\begin{equation}
U_{\rm target} = K_2 U_J(t) K_1,
\label{eq:decomp}
\end{equation}
where $K_1, K_2 \in K$ and 
\begin{equation}
U_J (t_i) = e^{-(i \pi J/2)( t_1 \sigma_x \otimes \sigma_x
+t_2 \sigma_y \otimes \sigma_y+t_3 \sigma_z \otimes \sigma_z)}.
\label{eq:uj}
\end{equation}
The total execution time is then given by $T=\sum_{i=1}^3 t_i$.
The group $SU(4)$ is 15-dimensional,
while each of the $K_i$ has six parameters 
and $U_J(t)$ has three, thus $2 \times 6 + 3 = 15$ parameters in total.
It follows from Eqs.~(\ref{eq:1qubit}) and (\ref{eq:2qubit}) that
the Hamiltonian (\ref{eq:ham}) contains the necessary and sufficient number of
control parameters to construct an arbitrary quantum gate.

\section{Time Optimal Solutions of Grover's Algorithm}

\subsection{Grover's Algorithm}

The solution of the decomposition (\ref{eq:decomp}) is difficult to
find and we have to resort to numerical methods in general.
As an example, we work out in the present section Grover's seach algorithm
\cite{ref:16} in the framework outlined in the previous section. 
Grover's algorithm for a two-qubit case is implemented by
unitary matrices of the form
\begin{eqnarray}
U_{00} &=& \left( \begin{array}{cccc}
-1&0&0&0\\
0&0&1&0\\
0&1&0&0\\
0&0&0&1
\end{array} \right)\label{eq:00}\\
U_{01}
&=& \left( \begin{array}{cccc}
0&0&1&0\\
-1&0&0&0\\
0&0&0&-1\\
0&-1&0&0
\end{array} \right)
\label{eq:01}\\
U_{10}&=& \left( \begin{array}{cccc}
0&1&0&0\\
0&0&0&-1\\
-1&0&0&0\\
0&0&-1&0
\end{array} \right)\label{eq:10}\\
U_{11}&=&\left( \begin{array}{cccc}
0&0&0&1\\
0&-1&0&0\\
0&0&-1&0\\
-1&0&0&0
\end{array} \right)\label{eq:11}
\end{eqnarray}
where $U_{ij}$ picks out the desired ``data'' $(ij)$ by operating on $|00
\rangle$ as
\begin{equation}
U_{ij}|00 \rangle = e^{i \alpha} | ij \rangle,\quad (i,j = 0, 1)
\end{equation}
where we have written explicitly the possible phase on the rhs \footnote{
The convensional pulse sequences \cite{ref:13} produce the matrices which
differ from (\ref{eq:00}) -- (\ref{eq:11}) in the {\it phase} of the matrix
elements, which does not bring about any observable difference.
}.

An interesting observation is that all the matrix elements are either
0 or $\pm 1$, which implies that $U_{ij}$ is not only in $SU(4)$ but also
in $SO(4)$. This further restricts the degrees of freedom of the matrix
since $SO(4)$ is six-dimensional.
In fact, our numerical results show that the decomposition 
(\ref{eq:decomp}) takes a highly constrained form.

\subsection{Optimization}

According to the presecription given in the previous section, we search
a pulse sequence of the form $U_{ij}= K_2 U_J K_1$. Instead of
solving this inverse problem, we define the equivalent variational
problem which is easy to solve, at least numerically. Let 
\begin{equation}
U[\gamma(t)] = K_2(\gamma_2)  U_J(t)
K_1(\gamma_1),
\end{equation}
where
\begin{equation}
K_i(\gamma_i) = e^{i (a_1 \sigma_x +b_1 \sigma_y + c_1 \sigma_z)}
\otimes e^{i (a_2 \sigma_x +b_2 \sigma_y + c_2 \sigma_z)}
\end{equation}
and $U_J(t)$ has been given in Eq.~(\ref{eq:uj}). Now we define the
penalty function $p[\gamma(t)]$ as
\begin{equation}
p[\gamma(t)] = \|U[\gamma(t)] -U_{\rm target}\|_{\rm F}, 
\end{equation}
where $\| A \|_{\rm F} = \sqrt{{\mathrm{tr}} A^{\dagger} A}$ is the
Frobenius
norm of a matrix $A$. Note that $p[\gamma(t)]$ is positive definite
and that its zeros are the absolute minima. Thus solving the inverse problem
is recast into a variational task for finding the absolute minima of 
$p[\gamma(t)]$. 

To find the absolute minima, we have generated 512 initial conditions
and searched for the optimal solutions with the polytope algorithm on a
parallel computer with 512 CPUs. To our surprise, the execution
time $T = \sum_{i=1}^3 t_i$ is discrete and assumes the values
\begin{equation}
T =(n + 1)/J, \quad n=0, 1, 2 \ldots
\end{equation}
for all four cases. 
The ambiguity $2\pi n$ corresponds
to the path leaving from $I_4$ and traversing the compact group
$SU(4)$ $n$ times before hitting the destination $U_{\rm target} K_2$.  
The result shows that Fig. 1 of \cite{ref:14} is misleading at least
in the present case: the distance
between the cosets $I_4 K_1$ and $U_{\rm target}K_2$ is {\it unique}
in the sense that the only degree of freedom left
is how many times the path traverses $SU(4)$ before arriving at the target.

The optimal execution time obtained here, however, is the same as
that for the conventional pulse sequence \cite{ref:13} in all cases. 
In other words, the conventional
pulse sequences are already time-optimal. For such a simple algorithm,
time-optimization may be carried out by inspection by experts. 
Therefore, we look for more complicated cases to demonstrate
the power of this method. Suppose
we would like to execute Grover's search algorithm with a room-temperature
liquid state NMR computer. The sample is in a thermal equilibrium state
and we will use the temporal averaging by cyclic permutations
of state populations to obtain a pseudopure
initial state. This is carried out by applying the unitary operators
\begin{equation}
U_{\rm{cp}} = \left(\begin{array}{cccc}
1&0&0&0\\
0&0&0&1\\
0&1&0&0\\
0&0&1&0
\end{array}
\right) = {\mathrm{CNOT}}_{12} {\mathrm{CNOT}}_{21}
\end{equation}
and
\begin{equation}
U_{\rm{cp}}^2
=\left(\begin{array}{cccc}
1&0&0&0\\
0&0&1&0\\
0&0&0&1\\
0&1&0&0
\end{array}
\right) = {\mathrm{CNOT}}_{21} {\mathrm{CNOT}}_{12}
\end{equation}
to the initial thermal state 
before $U_{ij}$ is executed. Here ${\mathrm{CNOT}}_{ij}$
stands for the CNOT gate with the control bit $i$ and 
the target bit $j$. 

We now present our search result for $U_{10}, U_{10} U_{\rm cp}$
and $U_{10} U_{\rm cp}^2$, the other ${ij}$ producing similar results.
\begin{itemize}
\item
$U_{10}$: An example of a typical time-optimal solution for $U_{10}$ is 
\begin{eqnarray}
K_1 &=& I_2 \otimes I_2\nonumber\\
U_J&=& e^{i (\pi/4)(\sigma_x \otimes \sigma_x-\sigma_y \otimes \sigma_y)}
\\
K_2 &=& e^{-i (\pi/4) \sigma_z} \otimes e^{i(\pi/2 \sqrt{2})
(\sigma_x + \sigma_y)}.\nonumber
\end{eqnarray}
The execution time is $T=1/J$, which simply
reproduces that for the conventional pulse sequence \cite{ref:13}. 
Therefore, the conventional pulse sequence is already optimized.
We will employ the conventional pulse sequence for $U_{10}$ in the
following. 
\item $U_{10} U_{\rm cp}$: For this case, an example of the time-optimal
pulse sequence is 
\begin{eqnarray}
K_1 &=& I_2 \otimes e^{-i (\pi/4) \sigma_x }
\nonumber\\
U_J&=& e^{-i (\pi/4)\sigma_z \otimes \sigma_z}\\
K_2 &=& e^{i (\pi/2) \sigma_y} \otimes  e^{ i(\pi/3 \sqrt{3})
(\sigma_x+ \sigma_y+ \sigma_z)}\nonumber\\
&=& e^{i (\pi/2) \sigma_y} \otimes \left[ e^{ i(\pi/4) \sigma_y}
e^{i(\pi/4)\sigma_x}\right],\nonumber
\end{eqnarray}
where the second line of $K_2$ shows the pulse sequence
made of the existing terms in the Hamiltonian.
It should be noted that this pulse sequence
requires the execution time $T = 1/2J$ in spite of an additional
gate $U_{\rm cp}$, which is composed of two CNOT gates and costs
$1/J$ of time to execute in the conventional pulse sequence. 
The execution time of the
time-optimal pulse sequence is 25\% of that for
the conventional pulse sequence. The execution time for the other
solutions takes discrete values $(n+1/2)/J,\ (n=0, 1, 2, \ldots)$.
This corresponds to a path traversing $SU(4)$ $n$ times
before arriving at the $U_{\rm target}$.
\item $U_{10} U_{\rm cp}^2$: An example of the time-optimal pulse
sequence is 
\begin{eqnarray}
K_1 &=& e^{-i (\pi/3 \sqrt{3})( \sigma_x +\sigma_y+\sigma_z)}
\otimes I_2
\nonumber\\
&=& \left[e^{-i(\pi/4) \sigma_x} e^{-i(\pi/4) \sigma_y}\right]\otimes
I_2 \nonumber\\
U_J&=& e^{-i (\pi/4)\sigma_z \otimes \sigma_z}\\
K_2 &=& e^{i (\pi/4) \sigma_x} \otimes I_2.\nonumber
\end{eqnarray}
Note again that the execution time of the gate for the time-optimal pulse 
sequence is 25\% of that for the conventional pulse sequence. In general,
the execution time is
$(n+1/2)/J,\ (n=0, 1, 2, \ldots)$ as in the above case.
\end{itemize}

The generators in the pulse sequences, that do not exist in 
the Hamiltonian (\ref{eq:ham}), are
rewritten in favor of the existing terms by making use of the
conjugate transformations (\ref{eq:1qubit}) and (\ref{eq:2qubit}).
The results are summarized in Table I.
Also shown in the Table are the results according to the 
conventional pulse sequence for the respective gate.

\begin{table*}[htb]
\begin{tabular}{|c|l|c|}
\hline
\multicolumn{3}{|c|}{Conventional pulse sequence}\\
\hline
Gate&Pulse sequence&Execution time\\
\hline
$U_{10}$&
\verb|1:                           -Y -(1/2J)-Ym-Xm-(1/2J)-Ym-Xm-|& 1/J\\
&
\verb|2:                           -Y -(1/2J)-Ym-X -(1/2J)-Ym-Xm-|& \\
\hline
$U_{10}U_{\rm cp}$&
\verb|1: -X -(1/2J)-X --------------Y -(1/2J)-Ym-Xm-(1/2J)-Ym-Xm-|& 2/J\\
&
\verb|2: --------------X -(1/2J)-X -Y -(1/2J)-Ym-X -(1/2J)-Ym-Xm-|&  \\
\hline
$U_{10}U_{\rm cp}^2$&
\verb|1: --------------X -(1/2J)-X -Y -(1/2J)-Ym-Xm-(1/2J)-Ym-Xm-|& 2/J\\
&
\verb|2: -X -(1/2J)-X --------------Y -(1/2J)-Ym-X -(1/2J)-Ym-Xm-|&  \\
\hline
\multicolumn{3}{|c|}{Time-optimal pulse sequence}\\
\hline
Gate&Pulse sequence&Execution time\\
\hline
$U_{10}$&
\verb|1:                       -X -(1/2J)-Xm-Ym-(1/2J)-Y -Pi(45)-|& 1/J\\
&
\verb|2:                       -X -(1/2J)-Xm-Y -(1/2J)-X -Ym    -|& \\ 
\hline
$U_{10}U_{\rm cp}$&
\verb|1:                                    -X -(1/2J)-Xm-Ym-    |& 1/2J\\
&
\verb|2:                                              -Ym-Ym-    |& \\
\hline
$U_{10}U_{\rm cp}^2$&
\verb|1:                                                         |& 1/2J\\
&
\verb|2:                                 -Y -X -(1/2J)-Xm-       |& \\
\hline
\end{tabular}
\caption{Control pulse sequences for Grover's algorithm,
which picks out the `file' $|10 \rangle$ 
starting from the pseudopure
state $|00\rangle$, which is obtained by cyclic permutations. 
The carbon nucleus is the first qubit while
the hydrogen nucleus is the second.
The upper Table shows the conventional pulse sequences \cite{ref:13}
required to execute Grover's algorithm with
the pseudopure state. The uppermost row shows
the pulse sequence for the gate $U_{10}$, while the second and the
third rows show those for $U_{10} U_{\rm cp}$ and $U_{10} U_{\rm cp}^2$,
respectively. Here
{\tt X} ({\tt Xm}) and {\tt Y} ({\tt Ym}) denote $\pi/2$ pulse 
along $x$ ($-x$) and $y$ ($-y$) axis,
respectively. 
The lower Table shows the time-optimal pulse sequences. The symbol
{\tt Pi(45)} 
denotes $\pi$-pulse along $(1,1,0)$ direction of the Bloch sphere. 
}
\label{table:ps}
\end{table*}

\subsection{Experiments}

In our experiments, we used 0.6 milliliter, 
200 millimolar sample of Carbon-13 labeled
chloroform (Cambridge Isotopes) in d-6 acetone \cite{ref:13}.
Data were taken at room 
temperature with a JEOL ECA-500 (the hydrogen Larmor frequency being
approximately 500~MHz) spectorometer \cite{ref:jeol}.
The measured coupling strength is $J=215.5$~Hz and
the transverse relaxation time $T_2$ is $\sim 7.5$~s for the
hydrogen nucleus while $\sim 0.30$~s for the carbon nucleus.  
The longitudinal relaxation time $T_1$ is measured to be
$\sim 20$~s for both nuclei.
The spin 1 and 2 in Table~\ref{table:ps} 
correspond to Carbon-13 and H, respectively.

\begin{figure}[h]
\begin{center}
\includegraphics[width=8cm]{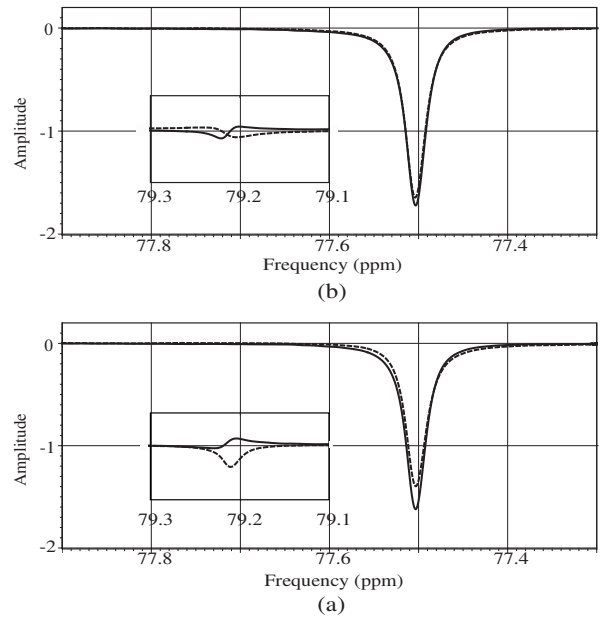}
\end{center}
\caption{
The spectra for the peak corresponding to $|10 \rangle$ state.
Negative signal amplitudes indicate that the carbon nucleus is in
the state
$|1 \rangle$, while appearance of the main signal at 
the frequency of 77.5~ppm (instead of 79.2~ppm) implies that
the hydrogen nucleus is in the
$|0 \rangle$ state. The insets show the signals in the vicinity of
79.2~ppm where a signal may appear if the hydrogen nulceus has
$|1 \rangle$ component. The scales in the insets 
are the same as in the main panels. 
(a) The upper panel shows the spectra obtained 
with conventional (dotted line) 
and optimized (solid line) pulse sequences. Each $\pi/2$-pulse duration is 
set to 25~$\mu$s. The main peak produced by the optimized
pulse sequences is slightly sharper than that of the conventional one.  
(b) The lower panel shows spectra 
with conventional (dotted line) 
and optimized (solid line) pulse sequences, in which the duration of the
$\pi/2$-pulse is now set to 250~$\mu$s. 
The signal produced by the optimized
pulse sequences is clearly better than that by the conventional ones.
Note also that unwanted signal in the inset is weaker for
the time-optimal pulse sequences.
}
\label{fig:s}
\end{figure}

Our experimental results are shown in Fig.~\ref{fig:s},
which shows the spectra corresponding to the $|1 0 \rangle$
state. The spectra in Fig.~~\ref{fig:s} (a) 
were obtained by using hard pulses whose duration is 25~$\mu$s
for $\pi/2$ pulses. We have intentionally introduced 
longer pulses with 250~$\mu$s duration to see the effect of imperfections,
see Fig.~\ref{fig:s} (b). In the first case, we can well ignore the time
evolution due to the J-coupling while pulses are applied since
the characterist time for the $J$-coupling is $\sim 1/J \sim 5$ms.
In the latter case, however, six 250~$\mu$s pulses amounts to the
total duration of 1.5ms, which is comparable to $1/J$ and we expect
that the difference in the number of pulses will manifest. 
Figure~\ref{fig:s} (b) clearly demonstrates 
that time-optimal pulse sequences produce sharper main peak compared to the 
conventional pulse sequences and less unwanted signal,
showing the superiority of our solutions.
We also expect that quantum algorithm accreleration should 
be effective to fight against decoherence. 

\section{Conclusions and Discussion}

In summary, we have demonstrated both theoretically and
experimentally that quantum algorithms 
may be accelerated if the unitary matrix realizing an
algorithm is directly implemented by manipulating the control
parameters in the Hamiltonian.
We have verified this by
implementing Grover's algorithm which picks out the ``file'' $|10 \rangle$
starting from the pseudopure state generated by cyclic permutations of
the state populations. We obtained the time-optimal 
pulse sequences and compared the results with those obtained by
the conventional pulse sequences. It turns out that the gate $U_{10}$
is already optimized in the conventional pulse sequence while the
gates $U_{10} U_{\rm cp}$ and $U_{10} U_{\rm cp}^2$ required for
the cyclic permutations are accelerated so that the execution time
is 25\% of that for the conventional pulse sequence in both cases.
The number of the pulses required for $U_{10} U_{\rm cp}$
($_{10} U_{\rm cp}^2$) is 4 (3) in the time-optimal pulses sequence, 
while it is 14 in both cases if the conventional pulse
sequences are employed. The smallness in the number of pulses required leads
to a higher-quality spectrum.

\section*{Acknowledgements}

We would like to thank Manabu Ishifune for sample preparation,
Toshie Minematsu for assistance in
NMR operations and Katsuo Asakura and Naoyuki Fujii of JEOL
for assistance in NMR pulse programming.
Parallel computing for the present work has been carried out with the
CP-PACS computer under the ``Large-scale Numerical Simulation Program''
of Center for Computational Physics, University of Tsukuba.
We would like to thank Martti Salomaa for drawing their attention to
Refs. \cite{ref:14, ref:15} and 
careful reading of the manuscript.
MN is grateful for partial support of a 
Grants-in-Aid for Scientific Research from Ministry of Education, Culture, 
Sports, Science and Technology (No.~13135215) and
Japan Society for Promotion of Science (JSPS) (No.~14540346).
ST would like to thank JSPS for partial support (No.~15540277).

\end{document}